\documentclass[nofootinbib,reprint,superscriptaddress,pdftex]{revtex4-1}
\usepackage{amsmath,amssymb}
\usepackage[dvips]{graphicx,color}
\usepackage[pdftex,colorlinks]{hyperref}
\numberwithin{equation}{section}
\begin{document}

\title{A maximum-entropy description of animal movement}
\author{Chris H. Fleming}
\affiliation{Smithsonian Conservation Biology Institute, Front Royal, Virginia}
\affiliation{Department of Biology, University of Maryland College Park}
\author{Yigit Subasi}
\affiliation{Department of Chemistry, University of Maryland College Park}
\author{Justin M. Calabrese}
\affiliation{Smithsonian Conservation Biology Institute, Front Royal, Virginia}
\date{\today}

\begin{abstract}
We introduce a class of maximum-entropy states that naturally includes within it all of the major continuous-time stochastic processes that have been applied to animal movement,
including Brownian motion, Ornstein--Uhlenbeck motion, integrated Ornstein--Uhlenbeck motion, a recently discovered hybrid of the previous models, and a new model that describes central-place foraging.
We are also able to predict a further hierarchy of new models that will emerge as data quality improves to better resolve the underlying continuity of animal movement.
Finally, we also show that Langevin equations must obey a fluctuation-dissipation theorem to generate processes that fall from this class of maximum-entropy distributions.
%Therefore, we extend the concept of a fluctuation-dissipation theorem beyond the context of thermodynamics and physics.
\end{abstract}

\maketitle

%%%%%%%
\section{Introduction}
Animals move continuously in time, with continuous velocities and accelerations, yet their locations are coarsely measured even by modern GPS technology.
Therefore, we are always limited by ignorance, as to the minutiae of detailed movements that occur at timescales our measuring apparatus cannot resolve.
To confront this problem, we derive a natural class of maximum-entropy states for stochastic processes that are assumed to be very continuous, but are only sampled at discrete time intervals.
The constraints with which we maximize entropy equate to understanding that a finite sampling frequency can only resolve the continuity of the sampled process to a finite degree.
As for all other behaviors of the process, we are guided by the principle of maximum entropy.

The class of maximum-entropy states we derive is found to include within it Brownian motion (BM) \citep{Horne07}, Ornstein--Uhlenbeck (OU) motion \citep{Dunn77,Blackwell03}, integrated OU motion \citep{Brillinger98,Johnson08,Gurarie11}, and a more general movement model that includes all of the previous models as limiting cases \citep{Variogram}.
The model derived in Ref.~\citep{Variogram} was motivated empirically, to fit the appearance of the autocorrelation structure in Mongolian gazelle telemetry data, and that it generalized previous continuous-time movement models was an apparent coincidence.
Here, we provide a theoretical framework that explains this coincident grouping of movement models in terms of continuity and entropy,
and we are able to predict a missing model within the same group that corresponds to central-place foraging theory.
We can also predict what models will become appropriate as GPS and battery technology improve to the point that more of the underlying continuity of animal movement is revealed.

Finally, we find that the multidimensional generalizations of these stochastic models obey a fluctuation-dissipation theorem (FDT).
In thermodynamic systems, fluctuations and dissipation are engendered by the same microscopic degrees of freedom, even though they are phenomenologically distinct.
%Because of their relation, 
%are intimately related by the \emph{fluctuation-dissipation theorem} (FDT). 
%At the heart of of non-equilibrium statistical mechanics, the FDT dates back to Einstein's work on atomic theory \cite{Einstein05} and Nyquist's work on electric conductivity \cite{Nyquist28}
As a simple example, for a damped mechanical system driven by thermal white noise, the Langevin equation is given by
\begin{align}
\mathbf{M} \, \ddot{\mathbf{x}}(t) + \hspace{-0.25em} \underbrace{ 2 \, \boldsymbol{\Gamma} \, \dot{\mathbf{x}}(t)}_{\mathrm{dissipation}} \hspace{-0.25em} - \mathbf{F}(\mathbf{x}(t)) = \hspace{-1em} \underbrace{ \boldsymbol{\xi}(t) }_{\mathrm{fluctuations}} \, ,
\end{align}
and the fluctuations and dissipation are be related by 
\begin{align}
\langle \boldsymbol{\xi}(t) \, \boldsymbol{\xi}(t')^\mathrm{T} \rangle &= \tilde{\boldsymbol{\sigma}}_{\!\xi\xi} \, \delta(t\!-\!t') \, , & %= 2 k_B T \, \delta(t\!-\!t') \, \boldsymbol{\Gamma} \, , \\
 \tilde{\boldsymbol{\sigma}}_{\!\xi\xi} &=  2 k_B T \, \boldsymbol{\Gamma} \, ,
\end{align}
where $\tilde{\boldsymbol{\sigma}}_{\!\xi\xi}$ is the spectral power of the fluctuations, $k_B$ is Boltzmann's constant and $T$ is the temperature of the surrounding environment.
The thermodynamic FDT is necessary for microscopic theories of stochastic processes to be consistent with macroscopic thermodynamics.
But we might imagine that dissipation coefficients and fluctuation autocorrelations are more generally unrelated---particularly in systems that have nothing to do with thermodynamics.
For the maximum-entropy distributions we explore here, we find that the fluctuations and dissipation are not necessarily proportional, but they must obey non-trivial commutation relations.
%, even though there are otherwise well-behaved Langevin equations that do not.

%%%%%%%%%
\section{Maximum-entropy states}
To constrain the degree of continuity in the underlying process, we will use the relationship between the continuity of the stochastic process and the continuity of its autocorrelation function.
Leaving everything else to ignorance, we do not privileged ourselves with information regarding the higher-order cumulants of the process,
and so upon constraining the mean $\boldsymbol{\mu}(t)$ and autocorrelation $\boldsymbol{\sigma}(t,t')$ functions, as defined in Eqs.~\eqref{eq:mean}-\eqref{eq:ACF},
the entropy per unit time functional is given by (App.~\ref{sec:Gaussian})
\begin{align}
h[\boldsymbol{\mu},\boldsymbol{\sigma}] &= \frac{1}{2} \int \!\! df \, \mathrm{tr} \log \tilde{\boldsymbol{\sigma}}(f) + \mathrm{constant} \, ,
\end{align}
in terms of the spectral-density function $\tilde{\boldsymbol{\sigma}}(f)$, defined by
\begin{align}
\boldsymbol{\sigma}(t,t') &= \int \!\! df \, e^{+2\pi \imath f (t-t')} \, \tilde{\boldsymbol{\sigma}}(f) \, ,
%(\log \boldsymbol{\sigma})(t,t') &= \int \!\! df \, e^{+2\pi \imath f (t-t')} \, \log \tilde{\boldsymbol{\sigma}}(f) \, .
\end{align}
for stationary autocorrelation, where $\boldsymbol{\sigma}(t,t') = \boldsymbol{\sigma}(t\!-\!t')$.
We consider only stationary autocorrelations, because they can be considered as the time average of non-stationary autocorrelations when estimating their parameters from a non-stationary process \citep{Likelihood2}.

%%%%%%
\subsection{Variance constraint}
As a simple example, we first consider a process with only its variance constrained to $\boldsymbol{\sigma}(0)$ and no further information:
\begin{align}
\left. \boldsymbol{\sigma}(\tau) \right|_{\tau=0} &= \boldsymbol{\sigma}(0) \, , &
\int \!\! df \, \tilde{\boldsymbol{\sigma}}(f)  &= \boldsymbol{\sigma}(0) \, ,
\end{align}
where the latter relation is conveniently expressed in the frequency domain.
The quantity to maximize, with Lagrange multiplier $\boldsymbol{\lambda}_0/2$, is given by
\begin{align}
m[\boldsymbol{\sigma}] &= h[\boldsymbol{\sigma}] + \frac{1}{2} \mathrm{tr} \, \boldsymbol{\lambda}_0 \left( \boldsymbol{\sigma}_0 - \int \!\! df \, \tilde{\boldsymbol{\sigma}}(f) \right) .
\end{align}
Using matrix derivatives \citep[][App.~B]{Equilibrium},
the Euler-Lagrange equations are then given by
\begin{align}
\frac{1}{2} \tilde{\boldsymbol{\sigma}}(f)^{-\mathrm{T}} &= \frac{1}{2} \boldsymbol{\lambda}_0^\mathrm{T} \, , &
\tilde{\boldsymbol{\sigma}}(f) &= \boldsymbol{\lambda}_0^{-1} \, , 
\end{align}
which implies that the spectral-density function is constant matrix.
I.e., the maximum-entropy process with variance $\boldsymbol{\sigma}(0)$ is a Markov process with variance $\boldsymbol{\sigma}(0)$.
The maximum-entropy process is not correlated in time without providing any further kinematic constraints.

%%%%%%
\subsection{Kinematic constraints and continuity}
The $k^\mathrm{th}$ derivative of $\mathbf{x}(t)$ has the autocorrelation function
\begin{align}
\frac{d^k}{dt^k} \frac{d^k}{dt'^k} \left\langle \left[ \mathbf{x}(t) \!-\! \boldsymbol{\mu}(t) \right] \left[ \mathbf{x}(t) \!-\! \boldsymbol{\mu}(t) \right]^\mathrm{T} \right\rangle &=
\frac{d^k}{dt^k} \frac{d^k}{dt'^k} \boldsymbol{\sigma}(t,t') \, .
\end{align}
Placing a constraint upon the $k^\mathrm{th}$ derivative of $\mathbf{x}(t)$ to have variance $\boldsymbol{\sigma}^{(k)}(0)$ takes the form
\begin{align}
\left. (-1)^k \frac{\partial^{2k}}{\partial \tau^{2k}} \boldsymbol{\sigma}(\tau) \right|_{\tau=0 }&= \boldsymbol{\sigma}^{(k)}(0) \, , \label{eq:kvar} \\
\int \!\! df \, (2\pi f)^2 \, \tilde{\boldsymbol{\sigma}}(f)  &= \boldsymbol{\sigma}^{(k)}(0) \, .
\end{align}
After maximizing entropy with these constraints, the spectral-density function is then given by
\begin{align}
\tilde{\boldsymbol{\sigma}}(f) &= \left[ \sum_{k=0}^K (2\pi f)^{2k} \boldsymbol{\lambda}_k \right]^{-1} , \label{eq:SDF}
\end{align}
when including kinematic constraints up to order $K$.

As any differentiable function is continuous, if a process has derivatives that always take finite values, then this process is always continuous.
Therefore, by placing kinematic constraints up to order $K$, we ensure that the process is continuous with $K-1$ continuous derivatives.
The $K^\mathrm{th}$ derivative of the process is not continuous, but is a well defined Markov process.

%%%%%%%
\subsubsection{K=1: OU \& BM motion}
As we have already shown, $K=0$ corresponds to uncorrelated motion of a particular variance.
$K=1$ corresponds to Ornstein--Uhlenbeck motion, which is a continuous process with autocorrelation function
\begin{align}
\sigma(\tau) &= \sigma(0) \, e^{-f |\tau|} \, ,
\end{align}
in one dimension.
This model describes Brownian motion within a spatial constraint, and ordinary Brownian motion is a limiting case for small $f$, where $\lim_{f \to 0} \sigma(0) \, f$ is the diffusion rate.

%%%%%%%
\subsubsection{K=2: OUF \& IOU motion}
$K=2$ includes within it OUF motion \cite{Variogram}, which is a continuous process with continuous velocities and autocorrelation function
\begin{align}
\sigma(\tau) &= \sigma(0) \, \frac{f_{+} \, e^{-f_{-} |\tau|} - f_{-} \, e^{-f_{+} |\tau|}}{f_{+}-f_{-}} \, . \label{eq:OUF}
\end{align}
This model describes bouts of autocorrelated velocity within a spatial constraint.
OUF motion reduces to OU motion in the limit $f_{+} \to \infty$ and to integrated OU motion in the limit $f_{-} \to 0$.
Integrated OU motion describes a process that is OU in velocities rather than positions and limits to Brownian motion when $f_{+} \to \infty$.
%The relationship among these four models is shown in Fig.~\ref{fig:phase}.
%\begin{figure}
%\centering
%\includegraphics[width=0.40\textwidth]{phase.pdf}
%\caption{\label{fig:phase} The parameter relationships among the four maximum-entropy models up to $K=2$,
%where $f_{-}$ corresponds to the first Lagrange multiplier that arose with $K=1$.}
%\end{figure}

%%%%%%%
\subsubsection{K=2: Central-place foraging}
Considering the general structure of Eq.~\eqref{eq:SDF},
there is one remaining model included in $K=2$ that has not previously been considered in the movement-ecology literature:
\begin{align}
\sigma(\tau) &= \sigma(0) \, e^{-f|\tau|} \left( \cos \omega \tau + \frac{f}{\omega} \sin \omega |\tau| \right) .
\end{align}
In this model there are periodic episodes of diffusion and anti-diffusion from and to the mean $\mu$.
The phenomenological behavior of this model is particularly relevant for describing central-place foraging,
where an animal has a nest or den at its mean location $\mu$ and periodically leaves to perform a random search for foraging patches.
This periodic motion stands in contrast to periodicities in the mean, such as migration,
where the animal cycles between its summering and wintering grounds.
The probability density of a central-place forager is unimodal, whereas the probability density of a migratory species is bi-modal.

%%%%%%%
\subsubsection{K=1: An excluded model}
It is also interesting to note what models are not included in this class.
For instance, the autocorrelation function
\begin{align}
\sigma(\tau) &= \sigma(0) \,  e^{-f |\tau|} \cos(\omega \tau) \, , \label{eq:cos}
\end{align}
does not have a spectral-density function consistent with Eq.~\eqref{eq:SDF} with any finite number of constraints,
even though this model is often considered as an oscillatory generalization of the OU process.

%%%%%%%
\subsection{Multi-variate Ornstein--Uhlenbeck motion}
Constraining the process up to its velocity results in the spectral-density function
\begin{align}
\tilde{\boldsymbol{\sigma}}(f) &= \left[ \boldsymbol{\lambda}_0 + (2\pi f)^2 \boldsymbol{\lambda}_2 \right]^{-1} , 
\end{align}
where both $\boldsymbol{\lambda}$ matrices must be positive definite for this to be a valid spectral-density function.
Factoring this expression, we have
\begin{align}
\tilde{\boldsymbol{\sigma}}(f) &= \boldsymbol{\lambda}_2^{-\frac{1}{2}}  \left[ \mathbf{F}^2 + (2\pi f)^2 \right]^{-1} \boldsymbol{\lambda}_2^{-\frac{1}{2}} \,  , \label{eq:OU1}
\end{align}
where $\mathbf{F}^2 = \boldsymbol{\lambda}_2^{-1/2} \boldsymbol{\lambda}_0 \, \boldsymbol{\lambda}_2^{-1/2}$
must then be a positive-definite matrix of square frequencies.
Fourier transforming back into the time domain, we have the autocorrelation function
\begin{align}
\boldsymbol{\sigma}(\tau) &= \boldsymbol{\lambda}_2^{-\frac{1}{2}} \frac{ e^{-|\tau| \mathbf{F}} }{2\, \mathbf{F}} \boldsymbol{\lambda}_2^{-\frac{1}{2}} \, . \label{eq:OU}
\end{align}
This describes a multivariate Ornstein--Uhlenbeck process of various dissipation rates and variance $\mathbf{I}$ that is linearly transformed to have variance
\begin{align}
\boldsymbol{\sigma}(0) &= \frac{1}{2} \, \boldsymbol{\lambda}_2^{-\frac{1}{2}} \, \mathbf{F}^{-1} \, \boldsymbol{\lambda}_2^{-\frac{1}{2}} \, . \label{eq:var}
\end{align}

%%%%%%%%
\section{Fluctuation-dissipation theorem}

\subsection{Ornstein--Uhlenbeck theorem}
To compare with Eq.~\eqref{eq:OU} and without loss of generality, we will consider the Langevin equation of a multivariate, mean-zero OU process $\mathbf{x}(t)$,
which represents the difference between the animal's location and its mean:
\begin{align}
\dot{\mathbf{x}}(t) &= \underbrace{ -\boldsymbol{\Gamma} \, \mathbf{x}(t) }_\mathrm{dissipation} \; + \; \hspace{-1em} \underbrace{ \boldsymbol{\xi}(t) }_\mathrm{fluctuations} \, ,
\end{align}
where $\boldsymbol{\xi}(t)$ is a Markov process with autocorrelation function 
\begin{align}
\langle \boldsymbol{\xi}(t) \, \boldsymbol{\xi}(t')^\mathrm{T} \rangle &= \tilde{\boldsymbol{\sigma}}_{\!\xi\xi} \, \delta( t\!-\!t') \, .
\end{align}
Note that $\tilde{\boldsymbol{\sigma}}_{\!\xi\xi}$ must be positive definite and real, and therefore it is symmetric.
Standardizing our Langevin equation so that the fluctuations have unit spectral density, we have
\begin{align}
\dot{\mathbf{y}}(t) &=  -\mathbf{G} \, \mathbf{y}(t) + \mathbf{u}(t) \, ,
\end{align}
in terms of the transformed variables
\begin{align}
\mathbf{y}(t) &= \tilde{\boldsymbol{\sigma}}_{\!\xi\xi}^{-\frac{1}{2}} \, \mathbf{x}(t) \, , &
\mathbf{u}(t) &= \tilde{\boldsymbol{\sigma}}_{\!\xi\xi}^{-\frac{1}{2}} \, \boldsymbol{\xi}(t) \, , \\
\mathbf{G} &= \tilde{\boldsymbol{\sigma}}_{\!\xi\xi}^{-\frac{1}{2}} \, \boldsymbol{\Gamma} \, \tilde{\boldsymbol{\sigma}}_{\!\xi\xi}^{+\frac{1}{2}} \, . \label{eq:G}
\end{align}
The dissipation matrices $\boldsymbol{\Gamma}$ and $\mathbf{G}$ are related by a similarity transform and therefore they share the same eigen-values, but in general they will not share the same symmetries.
Transforming to the frequency domain, we have
\begin{align}
2\pi\imath f \, \tilde{\mathbf{y}}(f) &= -\mathbf{G} \, \tilde{\mathbf{y}}(f) + \tilde{\mathbf{u}}(f) \, , \\
\tilde{\mathbf{y}}(f) &= \left[ 2\pi\imath f + \mathbf{G} \right]^{-1} \tilde{\mathbf{u}}(f) \, ,
\end{align}
and with this the spectral-density function is given by
\begin{align}
\tilde{\boldsymbol{\sigma}}(f) &= \langle \tilde{\mathbf{x}}(f) \, \tilde{\mathbf{x}}(f)^{\dagger} \rangle 
= \tilde{\boldsymbol{\sigma}}_{\!\xi\xi}^{+\frac{1}{2}} \, \langle \tilde{\mathbf{y}}(f) \, \tilde{\mathbf{y}}(f)^{\dagger} \rangle  \, \tilde{\boldsymbol{\sigma}}_{\!\xi\xi}^{+\frac{1}{2}} \, , \\
&= \tilde{\boldsymbol{\sigma}}_{\!\xi\xi}^{+\frac{1}{2}} \left[ \mathbf{G} + 2\pi \imath f  \right]^{-1} \left[ \mathbf{G} + 2\pi \imath f  \right]^{-\dagger} \tilde{\boldsymbol{\sigma}}_{\!\xi\xi}^{+\frac{1}{2}} \, .
%&= \tilde{\boldsymbol{\sigma}}_{\!\xi\xi} \left[ \boldsymbol{\Gamma}^2 + (2\pi f)^2  \right]^{-1} .
\end{align}
From Eq.~\eqref{eq:OU1}, if this is to represent a maximum-entropy state, then we must have 
\begin{align}
\tilde{\boldsymbol{\sigma}}_{\!\xi\xi} &= \boldsymbol{\lambda}_2^{-1} \, , &
\mathbf{G}  \, \mathbf{G}^\mathrm{T} &= \mathbf{F}^2 \, , &
\mathbf{G} &= \mathbf{G}^\mathrm{T} \, .
\end{align}
This final symmetry, applied to Eq.~\eqref{eq:G}, implies that the dissipation matrix and autocorrelation matrix must commute in the sense of
\begin{align}
\left[ \boldsymbol{\Gamma} , \tilde{\boldsymbol{\sigma}}_{\!\xi\xi} \right]_\mathrm{T} &= \boldsymbol{\Gamma} \, \tilde{\boldsymbol{\sigma}}_{\!\xi\xi} - \tilde{\boldsymbol{\sigma}}_{\!\xi\xi} \, \boldsymbol{\Gamma}^\mathrm{T} = \mathbf{0} \, ,
\end{align}
which reduces to ordinary commutation if $\boldsymbol{\Gamma}$ is symmetric.
We refer to this relation as comprising the Ornstein--Uhlenbeck fluctuation-dissipation theorem.
This FDT is more general (and weaker) than the thermodynamic relation, where the two matrices are strictly proportional.

%%%%%%%%
\subsection{General theorem}
The analogous Langevin equation for a continuous process $\mathbf{x}(t)$ with mean zero and $K-1$ continuous derivatives is given by
\begin{align}
\left[\frac{d}{dt}\right]^K \mathbf{x}(t) + \sum_{k=1}^{K} \boldsymbol{\Gamma}_k \left[ \frac{d}{dt} \right]^{K-k} \mathbf{x}(t) &= \boldsymbol{\xi}(t) \, .
\end{align}
By a similar procedure we have the transformed solutions
\begin{align}
\tilde{\boldsymbol{y}}(f) &= \left[ (2\imath\pi f)^K + \sum_{k=1}^{K} (2\imath\pi f)^{K-k} \mathbf{G}_k \right]^{-1} \tilde{\boldsymbol{u}}(f) \, ,
\end{align}
and for the spectral-density function to take the form \eqref{eq:SDF}, we must have the transformed commutation relations
\begin{align}
\mathbf{G}_1 &= \mathbf{G}_1^\mathrm{T} \, , &
\mathbf{G}_{k}^\mathrm{T} \, \mathbf{G}_{k+1} &= \mathbf{G}_{k+1}^\mathrm{T} \, \mathbf{G}_{k} \, ,
\end{align}
which then implies the commutation relations
\begin{align}
\boldsymbol{\Gamma}_1 \, \tilde{\boldsymbol{\sigma}}_{\!\xi\xi}&= \tilde{\boldsymbol{\sigma}}_{\!\xi\xi} \, \boldsymbol{\Gamma}_1^\mathrm{T} \, , &
\boldsymbol{\Gamma}_{k} \, \tilde{\boldsymbol{\sigma}}_{\!\xi\xi}^{-1} \, \boldsymbol{\Gamma}_{k+1}^\mathrm{T}
&= \boldsymbol{\Gamma}_{k+1} \, \tilde{\boldsymbol{\sigma}}_{\!\xi\xi}^{-1} \, \boldsymbol{\Gamma}_{k}^\mathrm{T} \, .
\end{align}
%[IS THERE AN INTERPRETATION OF WHAT THIS IS DOING, OTHER THAN MAKING THE END RESULT LOOK NICER?]

%%%%%%%%
\section{Range-residence versus central-place foraging}
In one dimension the mean-zero Langevin equation for $K=2$ is given by
\begin{align}
\ddot{x}(t) + 2 \, f \, \dot{x}(t) + F^2 \, x(t) &= \xi(t) \, ,
\end{align}
which is the equation of motion of a simple, damped harmonic oscillator driven by white noise.
Central-place foraging corresponds to the under-damped regime with relaxation rate $f$ and foraging frequency $\omega$ parameters
\begin{align}
f^2 &> F^2 \, , &
%f &= \frac{1}{2} \Gamma_1 \, , &
\omega &= \sqrt{F^2 - f^2} \, ,
\end{align}
where $\omega$ determines the frequency with which foraging bouts occur and $f$ determines the amount of correlation between successive foraging bouts.
In central-location foraging, the animal periodically leaves its mean location to search for resource patches and returns.
Just as a thermodynamic environment sets the Lagrange multiplier $T$ to its temperature,
an animal's environment can determine the animal's foraging frequency $\omega$, which is often fixed to $2\pi/$day.
%[IF YOU SAMPLE CENTRAL-PLACE FORAGING VERY COARSELY, THEN WILL IT REDUCE TO A K=1 MODEL?]

For the range-resident OUF model, which corresponds to the over-damped regime, the two relaxation rates are given by
\begin{align}
f^2 &< F^2 \, , &
f_\pm &= f \pm \sqrt{f^2-F^2} ,
\end{align}
where the smaller $f_{-}$ roughly determines the amount of correlation in successive positions and the larger $f_{+}$ roughly determines the amount of correlation in successive velocities.
In range-resident motion, the animal exhibits autocorrelated velocities within a finite home range.
Specifically for Mongolian gazelles, it has been observed that $f_{-}$ is associated with the seasonal timescale \citep{NDVI},
and so this Lagrange multiplier is also likely set by the environment.

Given that these two phenomenologically unrelated movement strategies---range residence and central foraging---can be placed into different parameter regimes of the same model,
and given that the parameters of this model are likely set by the environment,
we put forth the hypothesis that these movement strategies are, in fact, biologically related.
In the range-resident case, $1/f_{+}$ is a short timescale that determines the range of individual ballistic movements
and $1/f_{-}$ is the long timescale it takes for the animal to traverse its home range.
$1/f_{+}$ may represent, for instance, the time it takes to move between resource patches.
This perspective breaks down when $1/f_{+}$ exceeds $1/f_{-}$, and, in fact, our maximum entropy model suggests that in this case a transition occurs from range-residence to central-place foraging.
The biological interpretation of this transition is that the distance between resource patches exceeds a threshold value relative to the nesting area,
causing movement behavior to switch from a continuum of foraging to periodic bouts of foraging.

%%%%%%%
\section{Discussion}
We have placed all of the major continuous-time animal movement models within a simple framework, under the guise of maximizing entropy.
Our constraints are very natural for animal location data, in that, animal movement is extremely continuous, yet location data are relatively coarse,
and so we develop a hierarchy of models whereupon an increasing degree of continuity can be modeled and all finer scale behaviors are conceded to ignorance.
There are some mathematical similarities to Burg's maximum-entropy states for discrete-time processes, where the autocorrelation function is constrained up to a fixed number of lags \cite{Burg75}.
Burg derived the entire class of discrete-time auto-regressive (AR) processes, while we derive a restricted class of continuous-time auto-regressive processes that obey a fluctuation-dissipation theorem.
Otherwise, in both cases, understanding the importance of these models---in the context of maximizing entropy---is novel and interesting.

The OUF movement model was previously observed in Mongolian gazelle tracking data and it was motivated from empirical grounds in \cite{Variogram} and confirmed by maximum likelihood \citep{Likelihood2}.
Here, we have provided the OUF model with a statistical-kinematical interpretation.
GPS location fixes were obtained with a frequency sufficient to resolve the continuity of the gazelles' velocities but not their accelerations.
As the gazelles exhibit no migratory behavior, which would be encoded in the mean function, the relevant maximum-entropy state is either range residence (OUF) or the central-place foraging model that we have newly derived here.
The vast distances between good resource patches in the Eastern Steppe of Mongolia are then likely to be what make the gazelle range resident, rather than central-place foragers.
Moreover, as GPS and battery technology improve, possibly by combining telemetry and accelerometer data, our theory predicts that we can increase the number of kinematic constraints $K$ to derive more suitable models.

A natural question that arises from this perspective regards how strong the analogy between our maximum-entropy states and thermodynamics might be.
In both cases, the entropy is maximized with respect to natural constraints that regard what we can reasonably measure;
in both cases there are Lagrange multipliers that are determined by the environment;
and finally, in both cases there is a fluctuation-dissipation theorem, though in our case it is comparatively weak.
In Brownian motion derived from Hamiltonian mechanics, there will always be a relationship between the fluctuations and dissipation,
even outside of the context of thermodynamics \citep{FDR},
and so we might ask if there is any unifying microscopic theory that generates the FDT here and what sort of interpretation it has.

\appendix
%%%%%%%
\section{Derivation of entropy functional}\label{sec:Gaussian}
Here we will show a result that is well known for multivariate random variables---if we constrain ourselves to the first two cumulants or moments of the stochastic process,
then the distribution that maximizes entropy is the normal distribution.
The entropy of a distribution $p$ is given by
\begin{align}
H[p] &= -\int \!\! \mathcal{D}\mathbf{x} \, p[\mathbf{x}] \log p[\mathbf{x}] \, ,
\end{align}
and we will maximize it under the constraints
\begin{align}
1 &= \int \!\! \mathcal{D}\mathbf{x} \, p[\mathbf{x}] \, , \\
\boldsymbol{\mu}(t) &= \int \!\! \mathcal{D}\mathbf{x} \, p[\mathbf{x}] \, \mathbf{x}(t) \, , \label{eq:mean} \\
\boldsymbol{\sigma}(t,t') &= \int \!\! \mathcal{D}\mathbf{x} \, p[\mathbf{x}] \, \left[ \mathbf{x}(t) \!-\! \boldsymbol{\mu}(t) \right] \left[ \mathbf{x}(t') \!-\! \boldsymbol{\mu}(t') \right]^\mathrm{T} \, , \label{eq:ACF}
\end{align}
which is equivalent to maximizing 
\begin{widetext}
\begin{align}
M[p] &= H[p]
+ \lambda_0 \left( 1- \int \!\! \mathcal{D}\mathbf{x} \, p[\mathbf{x}] \right)
+ \int \!\! dt \, \boldsymbol{\lambda}_1(t)^\mathrm{T} \left( \boldsymbol{\mu}(t) - \int \!\! \mathcal{D}\mathbf{x} \, p[\mathbf{x}] \, \mathbf{x}(t) \right) \nonumber \\
&\phantom{=} + \iint \!\! dt \, dt' \, \mathrm{tr} \, \boldsymbol{\lambda}_2(t,t') \left( \boldsymbol{\sigma}(t,t') - \int \!\! \mathcal{D}\mathbf{x} \, p[\mathbf{x}] \, \left[ \mathbf{x}(t) \!-\! \boldsymbol{\mu}(t) \right] \left[ \mathbf{x}(t') \!-\! \boldsymbol{\mu}(t') \right]^\mathrm{T} \right) .
\end{align}
\end{widetext}
where the $\lambda$ are Lagrange multipliers.
The Euler-Lagrange equations are then given by
\begin{widetext}
\begin{align}
\log p[\mathbf{x}]+ 1 &= \lambda_0
+ \int \!\! dt \, \boldsymbol{\lambda}_1(t)^\mathrm{T} \mathbf{x}(t)
+ \iint \!\! dt \, dt' \, \mathrm{tr} \, \boldsymbol{\lambda}_2(t,t') \left[ \mathbf{x}(t) \!-\! \boldsymbol{\mu}(t) \right] \left[ \mathbf{x}(t') \!-\! \boldsymbol{\mu}(t') \right]^\mathrm{T} , \\
p[\mathbf{x}] &= \exp\!\left( \lambda_0 -1 
+ \int \!\! dt \, \boldsymbol{\lambda}_1(t)^\mathrm{T} \mathbf{x}(t)
+ \iint \!\! dt \, dt' \, \mathrm{tr} \, \boldsymbol{\lambda}_2(t,t') \left[ \mathbf{x}(t) \!-\! \boldsymbol{\mu}(t) \right] \left[ \mathbf{x}(t') \!-\! \boldsymbol{\mu}(t') \right]^\mathrm{T}
\right) .
\end{align}
\end{widetext}
Choosing the Lagrange multipliers that satisfy our constraints, we finally have
\begin{align}
p[\mathbf{x}] &= \frac{1}{\sqrt{\det 2 \pi \boldsymbol{\sigma}}} e^{-\frac{1}{2} \iint dt \, dt' \left[ \mathbf{x}(t') \!-\! \boldsymbol{\mu}(t') \right]^\mathrm{T} \boldsymbol{\sigma}^{-1}(t,t') \left[ \mathbf{x}(t') \!-\! \boldsymbol{\mu}(t') \right]} \, .
\end{align}
which is the distribution of a Gaussian stochastic process.
\pagebreak
The entropy of a Gaussian stochastic process is then given by
\begin{widetext}
\begin{align}
H[\boldsymbol{\mu},\boldsymbol{\sigma}] =& \frac{1}{2} \int \!\! \mathcal{D}\mathbf{x} \, p[\mathbf{x}] \left( \log \det 2 \pi \boldsymbol{\sigma} + \iint \!\! dt \, dt' \left[ \mathbf{x}(t) \!-\! \boldsymbol{\mu}(t) \right]^\mathrm{T} \boldsymbol{\sigma}^{-1}(t,t') \left[ \mathbf{x}(t') \!-\! \boldsymbol{\mu}(t') \right] \right) \nonumber , \\
=& \frac{1}{2} \left( \int \!\! dt \, \mathrm{tr} \, (\log \boldsymbol{\sigma})(t,t) + \iint \!\! dt \, dt' \mathrm{tr} \, \mathbf{I} \, \delta(t \!-\! t') \right) + \mathrm{constant} \, , \\
=& \frac{1}{2} \int \!\! dt \, \mathrm{tr} \, (\log \boldsymbol{\sigma})(t,t) + \mathrm{constant} \, ,
\end{align}
\end{widetext}
which only depends on the autocorrelation function and not the mean.
The mean is deterministic and does carry with it any entropy.

Viewing the autocorrelation function $\boldsymbol{\sigma}(t,t')$ as a large, positive-definite matrix,
its eigen-decomposition is given by
\begin{align}
\boldsymbol{\sigma}(t,t') &= \int \!\! df \, \mathbf{U}(t,f) \, \tilde{\boldsymbol{\sigma}}(f) \, \mathbf{U}(t',f)^\dagger \, , \\
\delta(t \!-\! t') \, \mathbf{I} &= \int \!\! df \, \mathbf{U}(t,f) \, \mathbf{U}(t',f)^\dagger\, , \\
\delta(f \!-\! f') \, \mathbf{I} &= \int \!\! dt \, \mathbf{U}(t,f)^\dagger \, \mathbf{U}(t,f') \, ,
\end{align}
where for stationary autocorrelations $\mathbf{U}(f,t)$ is a harmonic function, $\tilde{\boldsymbol{\sigma}}(f)$ is the spectral-density function, and $f$ is their frequency. 
The entropy functional is then given by
\begin{align}
H[\boldsymbol{\sigma}] &= \frac{1}{2} \int \!\! dt \! \int \!\! df \, \mathrm{tr} \!\left[ \mathbf{U}(t,f) \, \log \tilde{\boldsymbol{\sigma}}(f) \, \mathbf{U}(t,f)^\dagger \right] , \\
&= \frac{1}{2} \int \!\! df \, \mathrm{tr} \log \tilde{\boldsymbol{\sigma}}(f) \int \!\! dt \, \mathbf{U}(t,f)^\dagger  \mathbf{U}(t,f) \, , \\
&= \frac{\delta(0)}{2} \int \!\! df \, \mathrm{tr} \log \tilde{\boldsymbol{\sigma}}(f) \, , 
\end{align}
to within a constant. This quantity is infinite, but equivalent for maximization, the average entropy per unit time is given by 
\begin{align}
h[\boldsymbol{\sigma}] &= \frac{1}{2} \int \!\! df \, \mathrm{tr} \log \tilde{\boldsymbol{\sigma}}(f) + \mathrm{constant} \, , 
\end{align}
which is also the instantaneous entropy rate for a stationary autocorrelation function.

%%%%%%%%%
%\subsection{Non-stationary variance constraint}
%
%\begin{align}
%m[\boldsymbol{\sigma}] &= h[\boldsymbol{\sigma}] + \int \!\! dt \, \mathrm{tr} \, \boldsymbol{\lambda}(t) \left[ \boldsymbol{\sigma}(t,t) - \int \!\! df \, \mathbf{U}(t,f) \, \tilde{\boldsymbol{\sigma}}(f) \, \mathbf{U}(t,f)^\dagger \right]
%\end{align}
%Now if you vary any non-stationarity, lets call it $\theta(t)$, since it can only appear in $\mathbf{U}$ and its inverse/adjoint matrix, you get
%\begin{align}
%\frac{\delta m}{\delta \theta(t)} &= -\frac{\delta}{\delta \theta(t)} \int \!\! dt \, \mathrm{tr} \, \boldsymbol{\lambda}(t) \int \!\! df \, \mathbf{U}(t,f) \, \tilde{\boldsymbol{\sigma}}(f) \, \mathbf{U}(t,f)^\dagger \, ,
%\end{align} 
%which has to vanish right?
%If the autocorrelation function is stationary then $U(t,f) = e^{+2\pi\imath f t}$ and $U(t,f)^\dagger = e^{-2\pi\imath f t}$, and this will always be satisfied trivially because all of the time-dependence cancels.
%This seems rather impossible to satisfy otherwise.
%
%There is no explicit time dependence in the entropy, only variances. So when you constrain time dependences, you get no contribution from the entropy.
%But is this an unfair way to cast the problem by stuffing all of the time dependence into the eigen-functions?
%Something doesn't seem quite right to me.

\pagebreak

%%%%%%%
\bibliography{bib}

\end{document}